\documentclass[aps,twocolumn,prl,floatfix,longbibliography,superscriptaddress,amssymb]{revtex4-1}

\usepackage{graphicx}
\usepackage{amsmath}
\usepackage{sidecap}
\usepackage{lipsum}
\usepackage{color}
\usepackage{enumitem}
\usepackage[bottom]{footmisc}
\usepackage[normalem]{ulem} 

\usepackage[a4paper,left=13.5mm,right=13.5mm,top=27mm,bottom=27mm]{geometry}

\newcommand{\omegaL}{\omega_{\text{L}}}

\begin{document}

\bibliographystyle{apsrev4-1}

\title{Collective Mode Interferences in Light-Matter Interactions}

\author{Robert J. Bettles}
\affiliation{Joint Quantum Center (JQC) Durham--Newcastle, Department of Physics, Durham University, South Road, Durham DH1 3LE, United Kingdom}
\affiliation{ICFO-Institut de Ciencies Fotoniques, The Barcelona Institute of Science and Technology, 08860 Castelldefels (Barcelona), Spain}
\author{Teodora Ilieva}
\affiliation{Joint Quantum Center (JQC) Durham--Newcastle, Department of Physics, Durham University, South Road, Durham DH1 3LE, United Kingdom}
\author{Hannes Busche}
\affiliation{Joint Quantum Center (JQC) Durham--Newcastle, Department of Physics, Durham University, South Road, Durham DH1 3LE, United Kingdom}
\affiliation{Department of Physics, Chemistry and Pharmacy, Physics@SDU, University of Southern Denmark, 5230 Odense M, Denmark}
\author{Paul Huillery}
\affiliation{Joint Quantum Center (JQC) Durham--Newcastle, Department of Physics, Durham University, South Road, Durham DH1 3LE, United Kingdom}
\affiliation{Laboratoire Pierre Aigrain, Ecole normale sup\'{e}rieure, PSL Research University, CNRS, Universit\'{e} Pierre et Marie Curie, Sorbonne Universit\'{e}s, Universit\'{e} Paris Diderot, Sorbonne Paris-Cit\'{e}, Paris, France}
\author{Simon W. Ball}
\affiliation{Joint Quantum Center (JQC) Durham--Newcastle, Department of Physics, Durham University, South Road, Durham DH1 3LE, United Kingdom}
\affiliation{Kavli Institute for Systems Neuroscience, Norwegian University of Science and Technology, Olav Kyrres gate 9, 7491 Trondheim, Norway}
\author{Nicholas L. R. Spong}
\affiliation{Joint Quantum Center (JQC) Durham--Newcastle, Department of Physics, Durham University, South Road, Durham DH1 3LE, United Kingdom}
\author{Charles S. Adams}\email{c.s.adams@durham.ac.uk}
\affiliation{Joint Quantum Center (JQC) Durham--Newcastle, Department of Physics, Durham University, South Road, Durham DH1 3LE, United Kingdom}

\date{\today}

\begin{abstract}
We present a theoretical and experimental analysis of transient optical properties of a dense cold atomic gas. After the rapid extinction of a weak coherent driving field (mean photon number $\sim 1.5$), a transient `flash' is observed. Surprisingly the decay of the `flash' is faster than the decay of the fastest superradiant mode of the system. We show that this `faster than superradiance decay' is expected due to the interference between collective eigenmodes that exhibit a range of frequency shifts away from the bare atomic transition. Experimental results confirm that the initial decay rate of the superradiant flash increases with optical depth, in agreement with the  numerical simulations for the experimental conditions.
\end{abstract}

\maketitle

The optical properties of an ensemble of light scatterers can be dramatically modified when the scatterers behave collectively rather than independently. Certain collective effects such as superradiance and subradiance are well known and have been observed in a wide variety of systems \cite{Guerin2015,Facchinetti2016a,Meir2014,Javanainen2017b,Peyrot2018,Guerin2015,Facchinetti2016a,Roof2016,Corman2017,Cottier2018,Guerin2017,Bromley2016,Guerin2016,Jenkins2016a,Jennewein2016,Jennewein2018,Keaveney2012,Meir2014,Javanainen2017b,Peyrot2018,Meir2014,Jenkins2013,Hopkins2013,Bradac2017,Rohlsberger2010,VanLoo2013,Mlynek2014,Svidzinsky2010}. The phenomenon of superradiance derives its origins from the 1954 paper by Dicke \cite{Dicke1954} who predicted that a fully excited ensemble of $N$ quantum emitters enclosed within a volume much smaller than the resonant wavelength, $\lambda$, can decay with a peak intensity enhanced by a factor of $N^2$, although in practice this enhancement is strongly suppressed by interactions \cite{Gross1982}.  
Collective behavior however can also occur even with just a single excitation and in extended samples with dimension larger than a wavelength.
For single excitation the field excites a superposition of many-body collective eigenmodes with decay rates between $0$ and $N\Gamma_0$, where $\Gamma_0$ is the single emitter decay rate, depending on the geometry. Interestingly, one can exploit geometry to engineer the collective response and enhance the light-matter coupling for applications such as photon memories and gates \cite{Bettles2016,Shahmoon2017,Jenkins2013,Facchinetti2016a,Manzoni2017,Scuri2018,Mahmoodian2018}.

For large ensembles, each atom (or, more generally, emitter) can be modelled as a driven-dissipative electric dipole which interacts with every other dipole in the ensemble. The resulting recurrent scattering of each photon can modify the ensemble decay rate and linewidth as well as introducing lineshifts and many other phenomena. This can occur even when the inter-atomic spacing is greater than the wavelength, as is the case in this paper. 
Investigation of collective behavior in atomic clouds has a rich history \cite{Roof2016,Corman2017,Cottier2018,Guerin2017,Bromley2016,Guerin2016,Jenkins2016a,Jennewein2016,Jennewein2018, scully2009collective,Svidzinsky2010,Santo2020}.  
Other collective interference effects have been previously observed in many experiments such as quantum interference beats between different quantum states \cite{Haroche1976} or as a result of the relative motion of atoms \cite{Whiting2017}. However, the interference between collective eigenmodes predicted and observed in this letter has not previously been discussed.

In this paper, we investigate `single-photon superradiance' in a dense cold ensemble with dimension larger than the optical wavelength. We present a striking example where interference between simultaneously excited collective eigenmodes leads to faster than expected superradiant population dynamics and photon emission.
{Unlike previous works observing a so-called `super-flash' which decays faster than the lifetime due to motional dephasing \cite{Chalony2011,Kwong2014,Kwong2015}, in our experiment the atomic motion is  frozen over the timescale of the decay and the observed speed-up is due to collective interference effects.}
In addition, we observe that perhaps counter-intuitively, even for a simple system of two-level emitters, the resonance linewidth, $\Delta\omega$, is not trivially related to the collective emission time scale, $\tau$, $\Delta\omega \neq 1/\tau$. By careful simulations of the many-body collective response we show how these results arise from interference between collective modes. Since collective mode interference is, in fact, common in most  many-body light-matter interactions and important with regard to many applications, for example, in the stability of atomic clocks \cite{Bromley2016}, a deeper understanding of its significance is vital to advances in quantum technology.

The paper is organised as follows: First we outline the theoretical model of collective eigenmodes in an atomic ensemble. Next, we describe the experiment and present data on the measurement of the flash decay rates as a function of the optical depth of the medium. Finally, we show that the faster than expected superradiant decay is predicted by the coupled collective mode model and arises due to collective mode interference.

Both numerically and experimentally, we investigate the optical signal collected by a single mode fibre downstream of a cloud of cold atoms driven by a tightly focused weak laser beam (probe, $1/\rm{e}^2$-waist radius $1.8\lambda$) resonant with a two-level electric dipole transition.  After the laser is switched on, the atomic dipoles are driven into a collective steady state.  The emission dynamics are observed in the laser propagation direction by turning the laser off on a timescale faster than the resonant excitation lifetime $\tau_0=1/\Gamma_0$. The atomic polarization decays producing a bright fluorescent flash due to the sudden cancellation of the extinction \cite{Chalony2011,Kwong2014,Kwong2015}.
The atomic density is varied by varying the number of atoms in the trap.

To understand the resulting emission dynamics, we consider the simplest configuration first: a pair of two-level atoms. This will allow us to observe the behaviour of an individual eigenmode. For $N=2$ atoms, there are two collective eigenmodes; one `symmetric' where the dipoles oscillate in phase with each other, and one `anti-symmetric' where the dipoles oscillate $\pi$ out of phase with each other. The eigenmodes can be determined numerically by treating each atom as a driven classical electric dipole with fixed polarization (see Supplementary Material). Fig.\ \ref{fig:numericalFlash}\textbf{a} plots the decay rate $\Gamma_p$ and lineshift $\Delta_p$ of the symmetric (dark blue) and anti-symmetric (light blue) modes for varying atomic spacing $R$ after a laser pulse with $\sim 17\tau_0$ duration. As $R$ decreases, the magnitudes of both $\Gamma_p$ and $\Delta_p$ spiral outwards. Putting the dipoles side-by-side perpendicularly to the propagation direction of the laser beam means that both atoms see an identical driving field such that the overlap between the field vector and the anti-symmetric mode vector is zero, leaving just the symmetric mode. For $R=0.3\lambda$, the decay rate of the symmetric eigenmode is $1.55\Gamma_0$. The fluorescence in the forward direction at the end of the driving pulse is plotted in Fig.\ \ref{fig:numericalFlash}\textbf{c}, where $P_{\rm{tot}}$ is the total collected signal and $P_0$ is the signal during the steady state of the driving pulse in the absence of atoms.
As soon as the driving field is switched off, the signal decays exponentially as $\rm{e}^{-\Gamma t}$ with a constant decay rate of $\Gamma=1.55\Gamma_0$ (Fig.\ \ref{fig:numericalFlash}\textbf{e}, where we define the decay rate as $\Gamma=-\partial\log (P_{\rm{tot}}/P_0) / \partial t$), equal to the decay rate of the symmetric eigenmode which entirely determines the atomic dynamics. For the resonance lineshape of the steady state (inset, Fig.\ \ref{fig:numericalFlash}\textbf{c}), we also observe a Lorentzian with Full-Width--Half-Maximum (FWHM) $\Delta\omega=\Gamma=1.55\Gamma_0$ and linecentre at around zero detuning $\Delta=\omegaL-\omega_0\simeq0$, again as predicted by the single symmetric eigenmode.

\begin{figure}[t]
\includegraphics[]{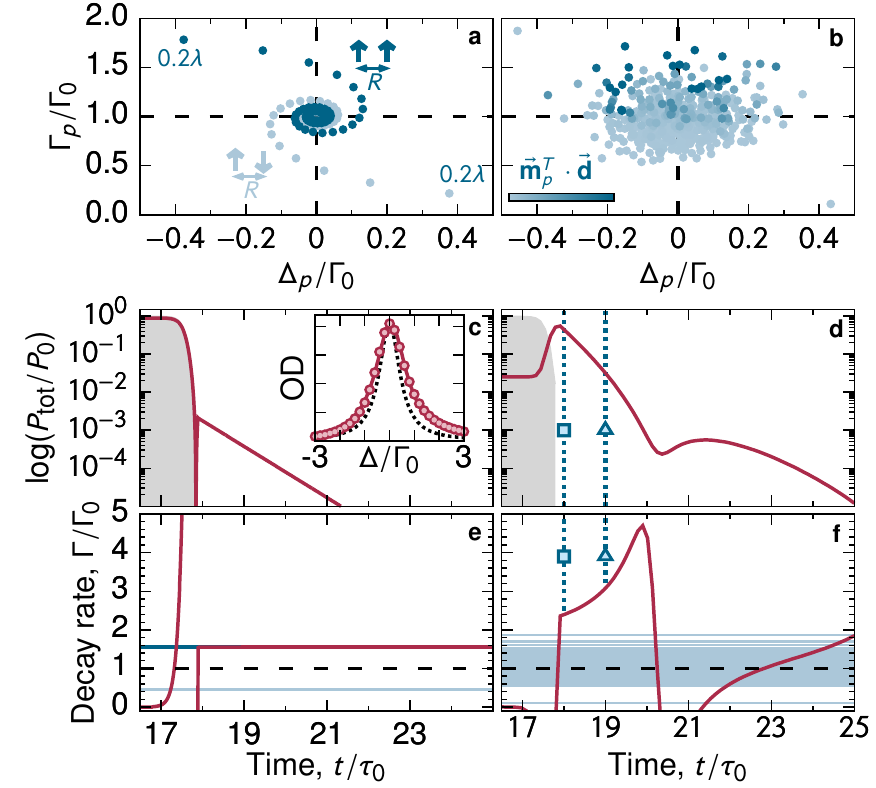}
\caption{{Flash decay rate and eigenmodes for different atomic ensembles.} 
\textbf{a}, \textbf{b}, The decay rates $\Gamma_{p}$ and frequency shifts $\Delta_{p}$ of the individual eigenmodes for $N=2$ with separation $R$ varying in steps of $0.05\lambda$ (\textbf{a}) and $N=400$ atoms for a single realisation (\textbf{b}). The shade of the dots is proportional to the coupling strength between the eigenvectors, $\vec{\mathbf{m}}_{\ell}$, and the polarisation vector, $\vec{\mathbf{d}}$.
\textbf{c}, \textbf{d}, Total relative optical power (red lines) and \textbf{e}, \textbf{f}, decay rate $\Gamma\equiv-\partial\log(P_{\rm{tot}}/P_0) / \partial t$ (red lines) of light collected in a waveguide along the probe propagation direction ($z$) following collective excitation by a weak probe pulse (grey shaded). \textbf{c}, \textbf{e}, For $N=2$ atoms separated in $x$ by $a=0.3\lambda$ and linearly polarized in $y$, the probe drives a single collective eigenmode producing a decay rate of $\Gamma=1.55\Gamma_0$ (dark blue line in \textbf{e}). The inset in \textbf{c} shows the steady state optical depth lineshape for the same pair of atoms (red circles) which are fitted to a Lorentzian with linewidth $\Delta\omega=1.55\Gamma_0$ (red solid line). The black dotted line shows the normalized optical depth due to a single atom. \textbf{d}, \textbf{f}, For {a single realization of} $N=400$ atoms in a cigar-shaped cloud (centre at $x=y=z=0$, see Fig.\ \ref{fig:setup}), a circularly polarised focused probe beam now couples to many eigenmodes, producing a time-varying decay rate which is initially much faster than any individual eigenmode decay rate (blue horizontal lines). The vertical blue dashed lines and markers indicate the position of decay rates in Fig.\ \ref{fig:decayRates}\textbf{a}. 
The waveguide surface is located at $z=250\lambda$ with radius $125\lambda$. }
\label{fig:numericalFlash}	
\end{figure}

\begin{figure}[t]
\includegraphics[]{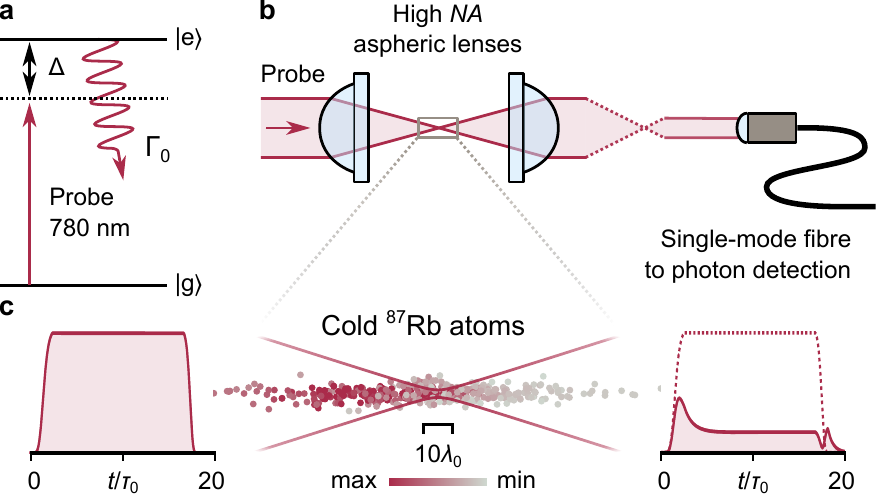}
\caption{{Overview of experiment}. \textbf{a} Scheme of relevant states in  $^{87}$Rb with $\lvert\textrm{g}\rangle=\lvert5S_{1/2},F=2,m_F=2\rangle$ and $\lvert\textrm{e}\rangle=\lvert5P_{3/2},F'=3,m_{F'}=3\rangle$.  \textbf{b} Experimental implementation. Using high-$NA$ aspheric lenses, a probe beam is tightly focussed ($1/e^2$- waist radius $w_{0}\approx 1\,\mathrm{\mu m}=1.28\lambda_0$) into a microscopic atomic ensemble confined in optical tweezers (not shown). Following re-collimation, light re-emitted in the original probe mode  is coupled into a single-mode fiber and detected using single photon detection modules. The red lines indicate the profile ($1/e^2$-width) of the original probe mode. \textbf{c} Temporal profiles of incoming probe pulse and emitted probe light ($\Delta=0$). The shade of the atoms indicates the simulated probability that an individual atom scatters a probe photon.}
\label{fig:setup}	
\end{figure}

The picture is significantly more complicated when there are many atoms which are arranged randomly. In Fig.\ \ref{fig:numericalFlash}\textbf{b},\textbf{d},\textbf{f} we consider $N=400$ atoms in a cigar shaped cloud 
{(Gaussian density distribution with standard deviations $\sigma_x=\sigma_y=1.5\,\mathrm{\mu m}$ (radial) and $\sigma_z=20\,\mathrm{\mu m}$ (axial), using the experimental parameters and $\lambda=780.24 \mathrm{nm}$)}
centred at the focus of a circularly polarised probe. In Fig.\ \ref{fig:numericalFlash}\textbf{b} we observe that for a single realisation, many of the $400$ possible eigenmodes now couple strongly to the driving field (dark blue markers), resulting in a time-dependent decay rate (\textbf{f}). \footnote{{The coupling strength between any individual eigenmode and the driving field is determined by the mode overlap between the two.}}
A changing decay rate is a natural consequence of faster superradiant modes decaying away, leaving only the slower subradiant modes \cite{Guerin2015}. However, simulations reveal an oscillatory behaviour and the effective decay rate can even temporarily become negative. Because of this oscillation, the initial decay rate ($\sim2.5\Gamma_0$) is actually faster than the decay of even the fastest single collective eigenmode ($\lesssim2\Gamma_0$) \footnote{Our claim is not that the decay is faster than $N\Gamma_0$ but rather the fastest possible eigenmode decay for an eigenmode of the considered configuration}\footnote{{For further analysis of the eigenmodes, see Fig.~S1 of [SuppMat]}}. This effect has also recently been seen in numerical simulations of a similar system \cite{Cottier2018}. The reason for this faster initial decay and oscillation is that shifted frequencies of the eigenmodes (dispersion in the eigenfrequencies) also significantly affect the dynamical behaviour. As we saw for the pair of atoms in Fig.\ \ref{fig:numericalFlash}\textbf{a},\textbf{c},\textbf{e}, each individual eigenmode is a mode of oscillation with its own distinct decay rate and resonance frequency. The total scattered field is then a sum of the emission from individual eigenmodes with shifted frequencies. These eigenfrequencies beat against each other and interfere. This interference can artificially change the transient decay rate and can thus result in decay rates significantly faster than any individual mode.
{The plots in Fig.~\ref{fig:numericalFlash} are for a single realization, although the behavior does not change significantly when averaging over many random realizations.}

We employ the experimental setup shown in Fig.\ \ref{fig:setup}\textbf{a}, details of which can be found elsewhere \cite{Busche2016,Busche2017}. In summary, a microscopic, cigar-shaped ensemble of a few thousand cold $^{87}$Rb atoms is confined in tightly focussed optical tweezers with estimated dimensions of $\sigma_x=\sigma_y=1.5\,\mathrm{\mu m}$ (radial) and $\sigma_z=20\,\mathrm{\mu m}$ (axial). The circularly polarised probe light (see Fig.~\ref{fig:setup}\textbf{b}) is tightly focussed into the ensemble ($1/\text{e}^2$-waist radius $w_{0,\mathrm{trap}}\approx 1\,\mathrm{\mu m}=1.28\lambda$) with detuning  $\Delta$ from the $\lvert5S_{1/2},F=2\rangle\rightarrow \lvert5P_{3/2},F'=3\rangle$ transition at $\lambda=780.24\,\mathrm{nm}$ with a natural linewidth of $\Gamma_0/2\pi=6.601\,\mathrm{MHz}$. Both probe and trap light are focussed using an aspheric lens (focal length $f=10\,\mathrm{mm}$, numerical aperture $NA\approx 0.5$). The light emitted from the ensemble in the forward direction is collected by a second, identical lens and detected behind a single-mode fibre which is aligned onto the mode of the incoming probe beam.

\begin{figure}[t]
\includegraphics[]{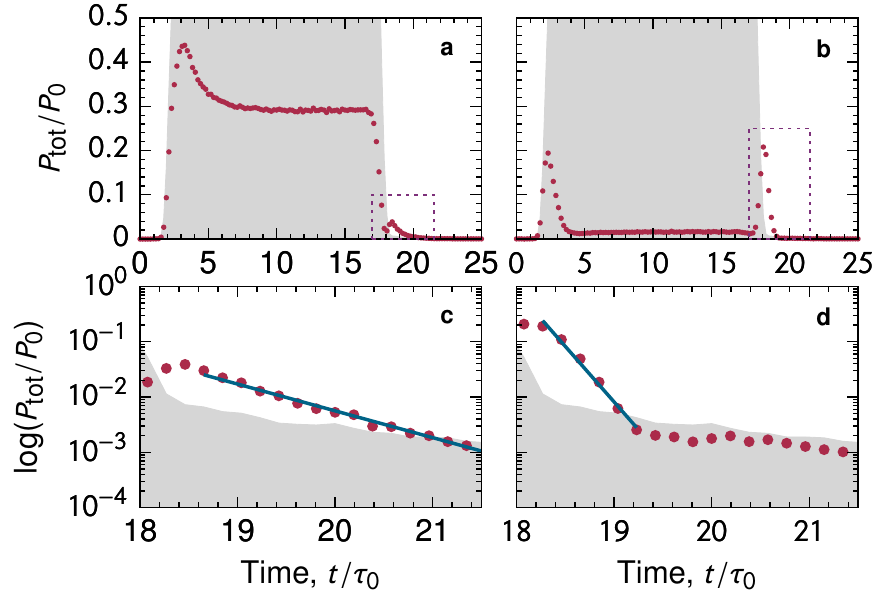}
\caption{{Experimental collective response of an atomic ensemble driven by an excitation pulse.} \textbf{a}, \textbf{b}, Total power collected by a single mode fibre. \textbf{c}, \textbf{d}, The decay rate of the flash $\Gamma$ is measured as the gradient of the log of the collected power (blue lines). The signal in the absence of atoms is indicated by the grey shaded region. \textbf{a}, \textbf{c}, Low atomic density produces low extinction and a slow decay rate ($\Gamma\simeq\Gamma_0$). \textbf{b}, \textbf{d}, High atomic density produces high extinction and a fast decay rate ($\Gamma\simeq4\Gamma_0$). 
}
\label{fig:experimentalFlash}	
\end{figure} 

\begin{figure}[t]
\includegraphics[]{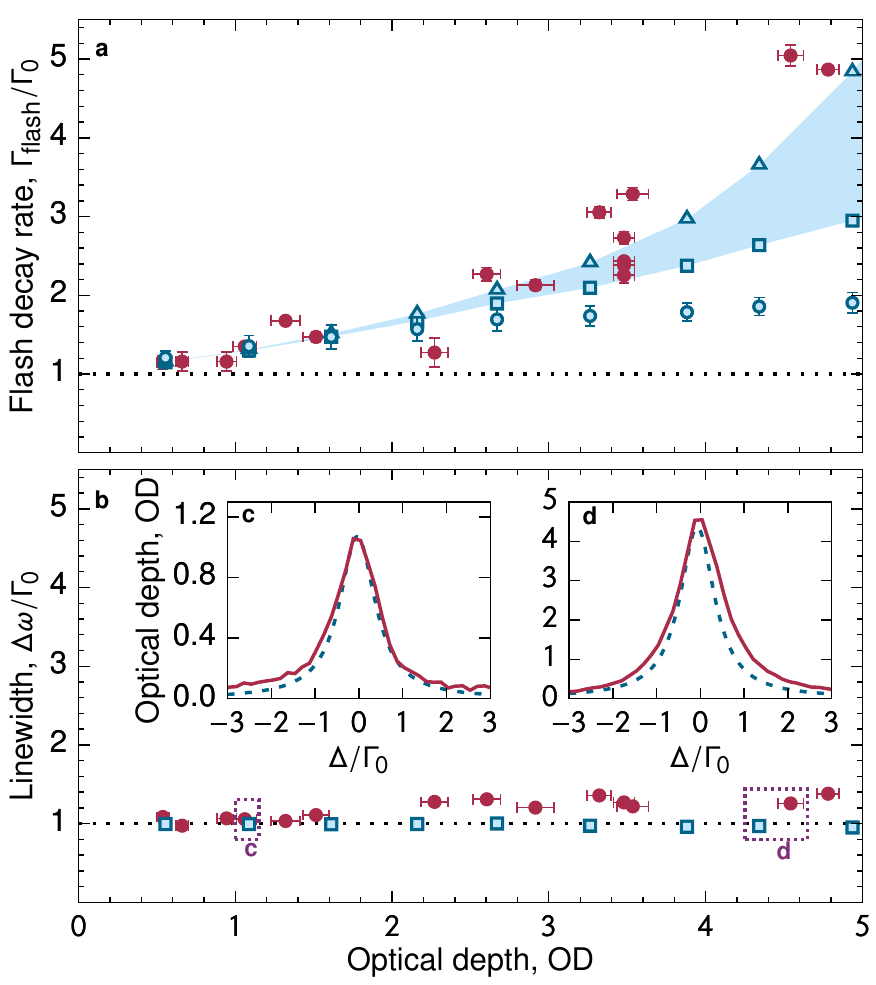}
\caption{{Experimental and numerical decay rates and linewidths.}
\textbf{a} Decay rate as a function of steady state optical depth. Experiment is shown with red circles and numerics with blue squares and triangles, which highlight the decay rate immediately after and $\tau_0$ after the pulse turn off respectively (see Fig.~\ref{fig:numericalFlash}\textbf{d},\textbf{f}). {The blue circles and errorbars plot the average and standard deviation of the maximum eigenvalue decay rates which for higher optical densities are consistently smaller than the numerical decay rates.}
\textbf{b} Full-width--half-maximum as a function of peak optical depth during steady state excitation for experiment (red circles) and numerics (blue squares). 
\textbf{c}, \textbf{d}, Lineshapes at low (\textbf{c}) and high (\textbf{d}) optical depth for both experiment (red solid lines) and numerics (blue dashed lines). 
{The numerical data is an average of 20 (a) and 100 (b,c,d) random realizations.}
}
\label{fig:decayRates}	
\end{figure}

Fig.\ \ref{fig:experimentalFlash} shows the experimental signal before, during, and after the optical driving pulse, for low (\textbf{a},\textbf{c}) and high (\textbf{b},\textbf{d}) atomic density. As in the numerics, we observe a flash after the pulse is switched off which decays exponentially for $1\tau_0$ to $3\tau_0$ before noise obscures the signal. Comparing the fitted decay rates with the numerical decay rates \footnote{{We quote the numerical decay rate immediately after and one lifetime after the pulse is switched off, as this corresponds to the shortest overall time window in any of our experiments before the signal became dominated by noise, Fig.~\ref{fig:experimentalFlash}\textbf{d}.}} in Fig.\ \ref{fig:decayRates}\textbf{a} we find that both experiment and theory demonstrate a clear positive trend of increasing initial decay rate with increasing optical depth $\rm{OD}$ [see [SuppMat] for details on determining and varying $\text{OD}=-\log (P_{\rm{tot}}/P_0)$],
{which are both in excess of even the fastest eigenmode decay rate (blue cirles).}
The increase with $\rm{OD}$ (and thus with increasing number density as the cloud dimensions remain similar) is further evidence of the superradiance being a collective effect \cite{Guerin2016}.
Contrary to numerical simulations, we do not observe any oscillation of the  decay rate during the specified time window. This could be a result of the higher atomic density required or the short accessible time window in the experiment.

Time--frequency correspondence implies that a change to the decay rate produces a change to the linewidth. For example, the decay rate and Full-Width--Half-Maximum (FWHM) linewidth for any given eigenmode are both $\Gamma_p$. However, despite the significant increase in the initial decay rate, we find that both experimentally and numerically there is relatively little increase in the linewidth. Initially, this seems counterintuitive and unphysical, but the linewidth--decay-rate correspondence does not take into account the large interferences we observe between the different eigenmode frequencies. Our decay rate window is defined only for the first one lifetime, where the superradiant modes are dominant. This picture is incomplete, as we have observed that within this time period the decay rate can vary greatly and be independent of linewidth. Investigation times are ultimately limited by the signal-to-noise ratio, defining the experimentally accessible region, and this must be accounted for when comparing decay rates and linewidths.

{There is good agreement between experiment and theory in the steady state lineshapes and overall trend of the linewidth and decay rate. One surprise is that we need far fewer atoms ($N=450$ for the maximum optical depth in Fig.~\ref{fig:decayRates}) in the simulations than we expect there to be in the experiment ($N\sim 5000$). This could imply that there are fewer atoms in the experiment than expected or that the numerics overestimates the true optical depth, which would also explain why the numerical decay rates and linewidths tend to be lower than the experimental ones. Other factors including beam misalignment, trap distortion or atomic dephazing may also have an effect.}
We are confident that we are in the weak driving regime, as the mean photon number per pulse is $\sim 1.5$ and also the measured decay rate is unaffected by varying the driving strength.

In summary, we have observed that interference between eigenmodes of collectively excited ensembles of optical dipoles can lead to counter-intuitive emission dynamics with decay rates seemingly faster than any individual superradiant eigenmode. 
The results presented imply a profound rethink of our understanding of light-matter interactions, in particular the relationship between resonance width and lifetime no longer necessarily holds in the presence of collective effects, especially when only considering dynamics over a short period of time.
At the same time, the good agreement between our simulations and experimental data provides a solid foundation to exploit collective mode engineering for applications in quantum technology.

\begin{acknowledgments}
R.J.B.\ thanks K.~M{\o}lmer and J.~Ruostekoski for illuminating discussions. 
This project has received funding from the European Union's Horizon 2020 research and innovation programme under grant agreement No. 640378 (FET-PROACT project ``RySQ''). This project has received funding from the European Union's Horizon 2020 research and innovation programme under grant agreement No. 660028 (Marie Sk\l{}odowska-Curie Individual Fellowship to P.H.). This project has received funding from the European Union's Horizon 2020 research and innovation programme under grant agreement No. 845218 (Marie Sk\l{}odowska-Curie Individual Fellowship to H.B.). 
We acknowledge funding from EPSRC through grant agreements EP/M014398/1 (``Rydberg soft matter''), EP/R002061/1 (``Atom-based Quantum Photonics''), EP/L023024/1 (``Cooperative quantum optics in dense thermal vapours''), 
EP/P012000/1 (``Solid State Superatoms''),
EP/R035482/1 (``Optical Clock Arrays for Quantum Metrology''),
EP/S015973/1 (``Microwave and Terahertz Field Sensing and Imaging using Rydberg Atoms''),
as well as, DSTL, and Durham University. 
R.J.B\ acknowledges financial support from the CELLEX-ICFO-MPQ Research Fellowships, MINECO Severo Ochoa Grant No. SEV 2015-0522, CERCA Programme/Generalitat de Catalunya, and Fundacio Privada Cellex.
The data presented in this letter are available at (DOI TO BE ADDED IN PROOF).
\end{acknowledgments}


%

\end{document}


\bibliographystyle{apsrev4-1}

\title{Supplemental Material to\\
``Collective Mode Interferences in Light--Matter Interactions''}

\maketitle

\noindent{\textbf{Experimental details.}}
We prepare the microscopic ensembles by loading laser cooled $^{87}$Rb atoms from a magneto-optical trap into a tightly confining optical dipole trap consisting of a single laser beam with wavelength $910\,\mathrm{nm}$ that is focussed to a $1/\text{e}^2$-waist radius of $\sim 4.5\,\mathrm{\mu m}$ and copropagates with the probe beam. 
Once the ensemble is prepared, we turn off the trap for $1.5\,\mathrm{\mu s}$ to remove any AC-Stark shifts due to the trap light and switch on the probe light for a pulse of $0.35\,\mathrm{\mu s}$. 
This procedure is repeated several thousand times before reloading the trap.
 
Instead of applying a separate optical pumping pulse for state preparation, the atoms are optically pumped into $\lvert g\rangle=\lvert5S_{1/2},F=2,m_F=2\rangle$ by the probe light during the first $1000$ to $1200$ repetitions. Consequently, these are not taken into account in our data analysis due to the rapid change of the optical depth during the pumping process.

In order to confine the atomic dynamics to just two internal energy levels, we apply an external magnetic field to Zeeman shift states with different $m_F$ and then optically pump into the maximal $m_F$ states in the ground and excited states, which are then coupled by probe light with circular polarization. 
To confirm that the two-level approximation is valid we varied the amount of time we waited after the optical pumping process before collecting any signal and found that other than the expected change in optical depth due to loss of number density this had no other effect. \\

\noindent{\textbf{Variation of the density/optical depth.}}
To measure linewidths and decay rates at different scatterer densities, we make use of the fact that the $\lvert 5S_{1/2},F=1\rangle$ level of the $^{87}$Rb ground state is not addressed by the probe light. We can thus control the fraction of atoms in $\lvert g\rangle$ by changing the duration of a repumping pulse that is applied on the $\lvert 5S_{1/2},F=1\rangle\rightarrow\lvert5P_{3/2},F=3\rangle$ transition after loading the dipole trap.

The resonant optical depth is determined by measuring the transmission $T(\Delta)=P_{\text{tot}}/P_0$ while the system is in the steady state (the period where the transmission of the probe pulse through the ensemble remains constant), for a range of probe detunings $\Delta$, and fitting the lineshape of the resulting spectrum $\text{OD}(\Delta)=-\log T(\Delta)$ to a Lorentzian including an offset, which is subsequently ignored and attributed to normalisation issues that arise from dark counts on the detectors. The resonant peak optical depth is then given by the amplitude. All errorbars stated for $\text{OD}$ correspond to the uncertainty of the fits.\\

\noindent{\textbf{Coupled dipole model.}} 
In the single-photon superradiance regime we assume there is only ever a single photon present in the cloud and as such can ignore quantum correlations and saturation effects, treating instead each atom as just a driven dissipative oscillating electric dipole. The response of dipole $j$, $\mathbf{d}_j$, is linear both to the driving laser field $\EL$ as well as the scattered fields from every other dipole,
\begin{equation}
\frac{\text{d}}{\text{d}t}{\mathbf{d}}_j = \left( \text{i} \Delta - \frac{\Gamma_0}{2} \right) \mathbf{d}_j + \text{i}\frac{|\mathcal{D}_0|^2}{\hbar}\left[\EL(\mathbf{r}_j) + \sum_{\ell\neq j} \mathsf{G}_{j\ell}\mathbf{d}_{\ell} \right],
\end{equation}
where $\Delta=\omegaL-\omega_0$ is the detuning of the laser field $\omegaL$ from the bare atomic resonance frequency $\omega_0$, $\Gamma_0$ is the spontaneous atomic decay rate for a single atom, $\mathcal{D}_0$ is the dipole matrix element of the atomic dipole transition, $\hbar$ is the reduced Planck constant, $\mathbf{r}_{j}$ is the position of dipole $j$, and $(\mathsf{G}_{j\ell}\mathbf{d}_{\ell})\equiv\mathbf{E}_{\ell}(\mathbf{r}_j)$ is the electric field at $\mathbf{r}_{j}$ scattered from dipole $\mathbf{d}_{\ell}$, 
\begin{align}
\mathbf{E}_{\ell}(\mathbf{r}_j) = & \frac{k^3}{4\pi\varepsilon_0}
\text{e}^{\text{i}kR}
\bigg\{ (\hat{\mathbf{R}}\times\mathbf{d}_{\ell})\times\hat{\mathbf{R}}\frac{1}{kR} 
\nonumber \\
& + \left[3\hat{\mathbf{R}} \left(\hat{\mathbf{R}}\cdot\mathbf{d}_{\ell}\right)-\mathbf{d}_{\ell}\right]
\left(\frac{1}{(kR)^3} - \frac{\text{i}}{(kR)^2}\right)
\bigg\},
\end{align}
where $\mathbf{R}\equiv\mathbf{r}_j-\mathbf{r}_{\ell}$ is the separation vector between atoms $j$ and $\ell$ with magnitude $R=|\mathbf{R}|$ and unit vector $\hat{\mathbf{R}}\equiv\mathbf{R}/R$.

The steady state where $(\text{d}/\text{d}t)\mathbf{d}_j=0$ can be represented as
\begin{equation}
\left(\frac{1}{\alpha} - \sum_{\ell\neq j}\mathsf{G}\right)\vec{\mathbf{d}} \equiv \mathsf{M}\vec{\mathbf{d}} = \vec{\mathbf{E}}_{\text{L}},
\end{equation}
where $\alpha = -(\mathcal{D}_0^2/\hbar)/[\Delta+\text{i}(\Gamma_0/2)]$ is the atomic polarisability, $\mathsf{G}$ is the matrix of all couplings $\{\mathsf{G}_{j\ell}\}$, and $\vec{\mathbf{d}}$ and $\vec{\mathbf{E}}_{\text{L}}$ are column vectors of all the dipole and laser field vectors. The coupling matrix $\mathsf{M}$ is not Hermitian but rather complex symmetric. This firstly means that the left and right eigenvectors are the transpose of each other rather than the conjugate transpose, meaning that under the standard vector dot product the eigenvectors $\mathbf{m}_p$ are non-orthogonal, i.e. $\vec{\mathbf{m}}_p^{\dag}\vec{\mathbf{m}}_q\neq\delta_{pq}$ but rather $\vec{\mathbf{m}}_p^{\text{T}}\vec{\mathbf{m}}_q=\delta_{pq}$. Secondly it means that the eigenvalues $\mu_p$ are complex with form 
\begin{equation}
\mu_p=-\frac{\hbar}{\mathcal{D}^2}\left[\left(\Delta-\Delta_p\right)+\text{i}\frac{\left(\Gamma_0+\Gamma_p\right)}{2}\right],
\end{equation}
which has the same form as the inverse atomic polarisability although with modified detuning $\Delta_p$ and linewidth $\Delta\omega=\Gamma_0+\Gamma_p$ which are proportional to the real and imaginary parts of the eigenvalues respectively. \\

\noindent{\textbf{Coupling into optical fibre.}} 
The experimental signal is proportional to the coupling of the total electric field into a single mode optical fibre. This coupling can be written as \cite{Aljunid2009,Jennewein2016}
\begin{equation}
\varepsilon = \int \left[ \mathbf{E}(\mathbf{r})\cdot \mathbf{g}^*(\mathbf{r})\right] \text{d}S,
\end{equation}
where the total field $\mathbf{E}(\mathbf{r})=\EL(\mathbf{r}) + \sum_j\mathbf{E}_j(\mathbf{r})$ is the sum of the driving field and total scattered field and $\mathbf{g}(\mathbf{r})$ is the mode of the single-mode fiber at position $\mathbf{r}$ integrated over an area $S$ perpendicular to the optical axis. The mode is matched to the laser field $\mathbf{g}\propto\EL$. The total power out is then proportional to 
\begin{equation}
P_{\text{tot}} \propto \langle |\varepsilon|^2  \rangle.
\end{equation}
The data presented in each figure is this power normalised by the equivalent signal in the absence of any atoms.

For the numerics we chose a focal length of $f=250\lambda$ and collection radius of $R=125\lambda$ which is slightly smaller than the experimental numerical aperture (0.5) although increasing the radius further beyond $125\lambda$ made negligible difference to the results. \\

\noindent{\textbf{Vector field propagation.}} The laser field is a Gaussian beam in the $\text{TEM}_{00}$ mode. However, the $1/e^2$ beam waist of $w_{0}\approx 1\,\mathrm{\mu m}=1.28\lambda_0$ is tightly focussed enough that the standard scalar model for paraxial propagation of a Gaussian beam produces incorrect results. One might assume the difference would only be a small correction, however we have found it to change both qualitatively and quantitatively the signal calculations. Therefore it is necessary instead to model the laser field propagation numerically. 

We employ the method used in \cite{Bettles2016,Tey2009,Adams2018}. The probe field starts off as a collimated Gaussian beam with $1/e^2$ beam radius $\wL$ incident on a focusing lens at position $z=-\zL$. As it passes through the lens the wavevector which starts off as $\mathbf{k}=k\,\hat{\mathbf{z}}$ is bent inwards towards the lens focus (the origin). If the field profile just before the lens was $\mathrm{i}E_{\text{L}} \mathop{\mathrm{e}^{-\rho^2/\wL^2}} \polvec_+$ (where $\polvec_{\pm}=(\hat{\mathbf{x}}{\pm}\mathrm{i}\hat{\mathbf{y}})/\sqrt{2}$ are two circular polarisations) then the field profile immediately after the (perfect thin) lens is 
\begin{widetext}
\begin{equation} \label{eq:ELvec}
\EL (\rho, \phi, z=-f) = \frac{E_{\mathrm{L}} \mathop{\mathrm{e}^{-\rho^2/w_{\mathrm{L}}^2}} }{\sqrt{|\cos\theta|}} \left( \frac{1 + \cos \theta}{2} \hat{\mathbf{\epsilon}}_+ + \frac{\sin\theta}{\sqrt{2}} \mathrm{e}^{\mathrm{i}\phi} \hat{\mathbf{z}} + \frac{ \cos \theta -1}{2} \mathrm{e}^{2\mathrm{i}\phi} \hat{\mathbf{\epsilon}}_- \right) 
 \mathop{\mathrm{\mathrm{exp}}} \left[-\mathrm{i} \left( k \sqrt{\rho^2 + f^2} - \pi/2 \right) \right],
\end{equation}
\end{widetext}
where $\rho^2 = x^2 + y^2$, $\phi = \tan^{-1}(y/x)$ and $\theta=\tan^{-1}(\rho/f)$ is the angle between the $-z$ axis and a point on the lens.

To propagate this field, it is first decomposed into an orthogonal set of modes $\EL = \sum_{\mu} \kappa_{\mu} \mathbf{E}_{\mu}$, where $\mu = (k_t,s,m)$, $k_t=\sqrt{k^2-k_z^2}$ is the transverse wavevector component, $s=\pm1$ is the helicity and $m$ is an angular momentum index. The expansion coefficients $\kappa_{\mu}$ can be calculated using 
\begin{widetext}
\begin{align} \label{eq:kappa_mu}
\kappa_{\mu} = \delta_{m1} \pi k_t \int_0^{\infty} \mathrm{d}\rho_{\textrm{L}}\, \rho_{\textrm{L}} \frac{1}{\sqrt{\cos\theta_{\textrm{L}}}} \Bigg\{ & \frac{sk+k_z}{k} \left( \frac{1+\cos\theta_{\textrm{L}}}{2} \right) J_0(k_t \,\rho_{\textrm{L}}) + \mathrm{i}\frac{\sqrt{2}k_t}{k} \left( \frac{\sin\theta_{\textrm{L}}}{\sqrt{2}} \right) J_1(k_t\,\rho_{\textrm{L}}) \nonumber\\ 
& + \frac{sk-k_z}{k} \left( \frac{\cos\theta_{\textrm{L}}-1}{2} \right) J_2(k_t\,\rho_{\textrm{L}}) \Bigg\} \mathrm{exp} \left[ -\mathrm{i} \left( k\sqrt{\rho^2_{\textrm{L}} + f^2} - \pi/2 \right) - \frac{\rho^2_{\textrm{L}}}{\wL^2} \right], 
\end{align}
\end{widetext}
where $J_m$ is the $m$th order Bessel function, $\rho_{\textrm{L}}$ is the radial position across the lens and $\theta_{\textrm{L}}=\tan^{-1}(\rho_{\textrm{L}}/f)$. The field components a distance $z$ from the lens focus in the $\pm$ and $z$ polarizations are then
\begin{widetext}
\begin{align} \label{eq:FullVecFieldPropagation}
E_+ (\rho,\phi,z) =& \EL \sum_{s=\pm1} \int^k_0 \mathrm{d}k_t \frac{1}{4\pi} \frac{s k + k_z}{k} J_0(k_t\rho) \mathop{\mathrm{e}^{\mathrm{i}k_z (z+f)}} \kappa_{\mu} , \nonumber \\
E_z (\rho,\phi,z) =& \EL \sum_{s=\pm1} \int^k_0 \mathrm{d}k_t (-\mathrm{i}) \frac{\sqrt{2}}{4\pi} \frac{k_t}{k} J_1(k_t\rho) \mathop{\mathrm{e}^{\mathrm{i}k_z (z+f)}} \mathop{\mathrm{e}^{\mathrm{i}\phi}} \kappa_{\mu} , \nonumber \\
E_- (\rho,\phi,z) =& \EL \sum_{s=\pm1} \int^k_0 \mathrm{d}k_t \frac{1}{4\pi} \frac{s k - k_z}{k} J_2(k_t\rho) \mathop{\mathrm{e}^{\mathrm{i}k_z (z+f)}} \mathop{\mathrm{e}^{2\mathrm{i}\phi}} \kappa_{\mu} .
\end{align}
\end{widetext}
The total field is then ${\EL = E_+ \polvec_+ + E_- \polvec_- + E_z \polvec_z}$. We calculate the field at the location of each atom as well as across the output lens, which recollimates the light using the inverse transform to (\ref{eq:ELvec}) and from which we can then calculate the signal coupled into the optical fibre.

\section{Eigenmode distribution}

\begin{figure}[t]
\includegraphics[]{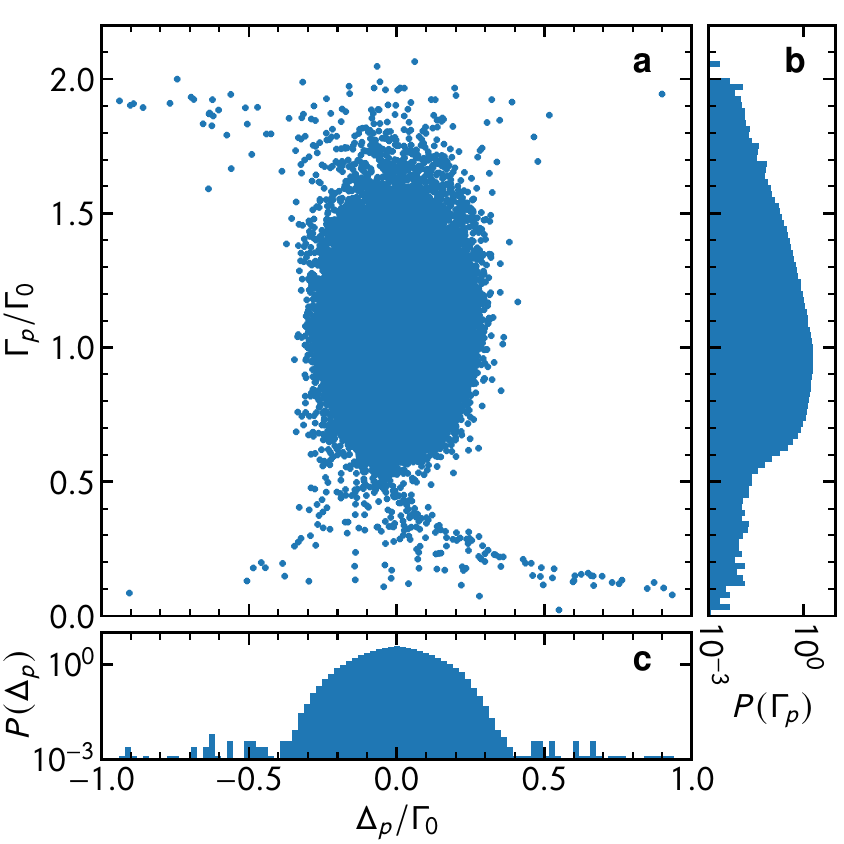}
\caption{{Eigenvalue distribution.
\textbf{a} Eigenmode shifts $\Delta_p$ and widths $\Gamma_p$ for an ensemble of $N=400$ interacting atoms using the same parameters as Fig.~1\textbf{b} from the main text, repeated $100$ times. \textbf{b},\textbf{c} Eigenvalue probability density $P$. 
}
}
\label{fig:Eigenmodes}	
\end{figure}

In Fig.~\ref{fig:Eigenmodes} we plot the eigenmodes for an ensemble of $N=400$ atoms as in Fig.~1\textbf{b} in the main text, except now showing $100$ random realizations rather than a single realization. This gives a more complete representation of the overall eigenmode distribution. Importantly, it shows that barely any eigenmode decay rates exceed $2\Gamma_0$ even for $100$ repetitions while the decay rate measured for the equivalent optical depth in Fig.~4\textbf{a} in the main text exceeds $\sim2.5\Gamma_0$. This excludes the possibility that the enhanced decay rate could be due to spurious large eigenmode decay rates.

%